\documentclass[11pt,letterpaper]{article}
\pdfoutput=1
\usepackage{jcappub}
\usepackage{bbm}
\usepackage{mathrsfs}
\usepackage{slashed}
\usepackage{caption}
\usepackage{epstopdf}
\usepackage[normalem]{ulem}
\usepackage[bottom]{footmisc}
\usepackage{subcaption}
\usepackage{bbold} 
\usepackage{titlesec}
\usepackage{threeparttable}
\usepackage{booktabs}
\usepackage{changepage}
\usepackage[utf8]{inputenc}
\usepackage{dsfont} 
\usepackage{grffile}
\usepackage{graphicx}  
\usepackage{dcolumn}   
\usepackage{bm}        
\usepackage{amssymb}   
\usepackage{setspace}
\usepackage{amsmath, amssymb, setspace}
\usepackage{array}
\usepackage{booktabs}
\usepackage{caption}
\usepackage{indentfirst}
\usepackage{float}
\usepackage{lmodern}
\usepackage{multirow}
\usepackage{soul}
\usepackage[normalem]{ulem}
\usepackage{braket}
\usepackage{comment}
\usepackage[draft]{pgf}
\usepackage{adjustbox} 
\usepackage{xspace} 

\newcommand{\DoBox}[1]{\begin{center}
\color{red}\fbox{
\begin{minipage}{0.9\textwidth}
\end{minipage}}
\end{center}}

\newcommand{\unit}[1]{\ensuremath{\mathrm{\,#1}}\xspace}

\newcommand{\GeV}{\unit{GeV}}

\newcommand{\cm}{\unit{cm}}

\newcommand{\kpc}{\unit{kpc}}

\newcommand{\Msolar}{\unit{M_\odot}}

\newcommand{\MBH}{\unit{M_{\mathrm{BH}}}}
\newcommand{\svu}{\mathrm {cm} ^{3}\;\mathrm {s} ^{-1}}
\newcommand{\sv}{\left\langle \sigma v \right\rangle}
\newcommand{\jf}{J_{\rm factor}}
\newcommand{\bb}{b\bar{b}}
\newcommand{\tautau}{\tau^{+}\tau^{-}}

\newlength{\myimageoversize}
\newsavebox{\myimage}

\usepackage[most]{tcolorbox}
\usepackage{empheq}

\tcbset{colback=white!10!white, colframe=black!50!black, 
        highlight math style= {enhanced, 
            colframe=black,colback=white!10!white,boxsep=0pt}
        }

\begin{document}
\title{\huge{Investigating the dark matter minispikes with the gamma-ray signal from the halo of M31}}

\author{Zi-Qing Xia$^{a}$\footnote{Corresponding author.},}
\author{Zhao-Qiang Shen$^{a}$\footnote{Corresponding author.},}
\author{Xu Pan$^{a,b}$,}
\author{Lei Feng$^{a,b,c}$\footnote{Corresponding author.},}
\author{Yi-Zhong Fan$^{a,b}$}

\emailAdd{xiazq@pmo.ac.cn}
\emailAdd{zqshen@pmo.ac.cn}
\emailAdd{xupan@pmo.ac.cn}
\emailAdd{fenglei@pmo.ac.cn}
\emailAdd{yzfan@pmo.ac.cn}

\affiliation{
$^a$Key Laboratory of DM and Space Astronomy, Purple Mountain
Observatory, Chinese Academy of Sciences, Nanjing 210033, China \\
$^b$School of Astronomy and Space Science, University of Science and Technology of China, Hefei, Anhui 230026, China\\
$^c$Joint Center for Particle, Nuclear Physics and Cosmology,
Nanjing University – Purple Mountain Observatory, Nanjing 210093, China\\
}

\date{\today}
\abstract{Recently, the evidence for gamma-ray emission has been found in the $Fermi$-LAT observation for the outer halo of Andromeda galaxy (M31).
The dark matter (DM) annihilation offers a possible explanation on the gamma-ray radiation.
In this work, we focus on the dark matter annihilation within minispikes around intermediate-mass black holes (IMBHs) with masses ranging from $100~\Msolar$ to $10^6~\Msolar$.
When the thermal annihilation relic cross section $\sv = 3 \times 10^{-26}~\svu$ is adopted, we conduct an investigation on the population of IMBHs in the spherical halo area of M31.
We find that there could be more than 65 IMBHs with masses of $ 100~\Msolar$ surrounded by the DM minispikes as the remnants of Population III stars in the M31 spherical halo, and it is almost impossible for the existence of minspikes around IMBHs with masses above $10^4~\Msolar$ which could be formed by the collapse of primordial cold gas, for both dark matter annihilation channels $\bb$ and $\tautau$.
The properties of dark matter have been further explored with the simulation of these two scenarios for IMBHs formation.}


\maketitle

\section{Introduction}

Dark matter (DM) is proposed to explain the missing mass which makes up approximately 85$\%$ of the matter in the current universe.
There are a large number of astrophysical observations establishing the presence of DM~\cite{1998TuckerApJ,2000CorbelliMNRAS}. 
But the nature of DM is still far from clear.
Among many hypothetical particle models, Weakly Interacting Massive Particles (WIMPs) are perceived as one of the most promising DM candidates~\cite{1996GerardPhR,2005BertonePhR,2005GianfrancoPhR,2010FengARAA}.
WIMPs could annihilate into standard model particles which can further produce gamma rays and cosmic rays. 
The present-day energy density of DM reveals that the cross section of WIMPs annihilation averaged over the velocity distribution through the s-wave process is about $\sv = 3 \times 10^{-26}~\svu$, called as the thermal relic cross section~\cite{2012SteigmanPhysRevD}.
The end products of WIMPs annihilation could be observed by the astronomical telescopes, which provides a feasible way to search for DM indirectly.

The observations of the gamma-ray sky have been widely used to indirectly detect DM~\cite{2016CharlesPhR}. The dwarf spherical galaxies~\cite{2011AckermannPRL,2014AckermannPhysRevD,2015AckermannPRL,2015HooperJCAP,2015GeringerPRL,2015LiPRD,2018LiPRD} and the Galactic center~\cite{2011HooperPLB,2013GordonPRD,2015ZhouPRD,2015CaloreJCAP,2016HuangJCAP,2016AjelloAPJ,2017AckermannApJGCE} are prevailing targets for WIMPs searches.
Especially for the dwarf galaxies, due to the high mass-to-light ratio, their observations have set the most stringent limits on WIMPs~\cite{2015AckermannPRL}.
Nearby galaxies are also regarded as interesting targets~\cite{Abdo2010A&A,Lenain2011A&A,2016LiJCAP,2017AckermannApJ,2017FuMNRAS,2019MauroPRD,2019KarwinApJ}.
M31 is the nearest large galaxies, its gamma-ray emission was first detected by the Fermi Large Area Telescope ($Fermi$-LAT) in Ref.~\cite{Abdo2010A&A}, and the corresponding DM constraints were then presented in Ref.~\cite{2016LiJCAP}.
Ref.~\cite{2019MauroPRD} has found that with the smooth DM spatial distribution, signals of DM particle interactions were insufficient to explain all the gamma-ray emission from M31 which was detected with the $0.4^\circ$ extension.

Interestingly, Ref.~\cite{2019KarwinApJ} separated M31 into three spherically symmetric components centered at the M31 (i.e. the inner galaxy, spherical halo and far outer halo) and reported a gamma-ray excess toward the halo region of M31.
The DM origin of the gamma-ray signal from the M31 spherical halo (M31SH) have been studied in Ref.~\cite{2021KarwinPhRvD,2021BurnsPhRvD}. 
Tab.~II in Ref.~\cite{2021KarwinPhRvD} has shown that in the case of the smooth Navarro-Frenk-White (NFW) density distribution in M31SH, the best-fit DM annihilation cross section $\sv = 787 \times 10^{-26}~\svu$ (with the best-fit DM mass $m_{\chi}=50~\GeV$) is obviously far above the thermal relic cross section $\sv = 3 \times 10^{-26}~\svu$ and has been excluded by the limits from the dwarf spherical galaxies observation.
Except for the annihilation cross section, the gamma-ray flux from the DM annihilation is also proportional to the square of the DM density.
It implies that with the thermal relic cross section, enhancements of the DM density (compared with the smooth NFW profile) are needed to yield the measured high gamma-ray flux toward the M31SH.
Substructures (or called subhalos) in the DM halos of both M31 and the Milky Way have been introduced to explain this issue in Ref.~\cite{2021KarwinPhRvD,2021BurnsPhRvD}.

In this paper, we propose another scheme to enhance the DM density in M31SH by using the minispikes around intermediate-mass black holes (IMBHs).
IMBHs from $100~\Msolar$ to $10^6~\Msolar$ are supposed to be the generic prediction of supermassive black holes and could originate from the collapses of Population III (Pop. III) stars or primordial cold gas in the early universe~\cite{2004MillerIJMPD,2005BertonePhRvD}.
Some of the ultraluminous X-ray sources are thought to be the IMBH candidates~\cite{2004SwartzApJS,2014PashamNature}.
If the IMBHs exist in the M31SH region, due to their adiabatic growth, the surrounding DM would form the minispike structure with high densities~\cite{2005ZhaoPhRvL}.
Due to the limited spatial resolution of $Fermi$-LAT, the DM minispikes could contribute the gamma-ray excess of M31SH as the unresolved point sources, providing a possible explanation on the high flux from the M31SH.
Since they were proposed, the minispikes around IMBHs have been widely studied for the DM indirect detection with the observations of the gamma-ray telescopes~\cite{2005ZhaoPhRvL,2007FornasaPhRvD,2008FornasaIJMPD,2008AharonianPhRvD,2009TaosoPhRvD,2009BringmannPhRvL,2009BertoneNJPh,2014AlmaPhRvD,2015WandersJCAP,2018LacroixApJ,2020Chengmnras,2021FortesJCAP} and the neutrino telescope~\cite{2006BertonePhRvD}.

Here we will investigate on the population of IMBHs in M31SH using the observed gamma-ray spectrum in the framework of the DM self-annihilation.
The paper is organized as follows: In Sec.~\ref{DMmodel}, we present the DM model used to calculate the gamma-ray flux from WIMPs annihilation in M31SH and introduce two proposed scenarios for IMBHs formed in the early universe.
In Sec.~\ref{sec:LPIMBHs}, we use the gamma-ray spectrum of M31SH to set limit on the population of IMBHs given the thermal relic cross section $\sv = 3 \times 10^{-26}~\svu$.
In Sec.~\ref{sec:discussion}, the $Scenario\ PopIII$ and $Scenario\ PCGas$ of IMBHs are discussed in detail for the M31SH to constrain the property of WIMPs. 
Sec.~\ref{sec:summary} summarizes our results and discussions.

\section{Dark matter model}\label{DMmodel}
\subsection{Gamma-rays from DM annihilation}
The WIMPs are expected to annihilate with each other and then produce gamma rays through many different annihilation channels (i.e. $\bb$ or $\tautau$).
The prompt gamma-ray flux from the DM annihilation can be given as a function of photon energy $E_{\gamma}$~\cite{2016CharlesPhR,2019MauroPRD}
\begin{equation}
\label{dPhide}
\frac{d\Phi}{dE_{\gamma}}(E_{\gamma})=  \frac{\left\langle \sigma v \right\rangle}{8 \pi  m_\chi^2} \frac{dN_\gamma}{dE_{\gamma}}\times \jf ,
\end{equation}
where $m_\chi$ denotes the rest mass of the DM particles, $\frac{dN_\gamma}{dE_{\gamma}}$ represents the differential photon production for the annihilation of a DM particle pair and can be obtained from the PPPC 4 DM ID\footnote{\url{http://www.marcocirelli.net/PPPC4DMID.html}}~\cite{2011PPPCJCAP}, $ \jf$ characterizes the spatial distribution of the DM and is the line-of-sight (l.o.s.) integral of the square of the DM density $\rho^2(r)$. The $ \jf$ within the region of interest (ROI) in one DM halo can be expressed as
\begin{equation}
\label{Jfactor}
\jf=\int_{\Delta\Omega}\int_{\text{l.o.s.}}\rho^2(\mathbf{r}(s))dsd\Omega,
\end{equation}
where $\Delta\Omega$ is the solid angle of ROI, $s$ is the distance along the line of sight and $r$ is the radial distance from the center of the halo.

\subsection{DM density distribution}
The DM density distribution in the region of M31SH can be divided into two types in our work: the Navarro-Frenk-White (NFW) profile and the Minispike (mnsp) profile.
The DM density in the main galactic halo of M31 is often approximated with a NFW profile~\cite{1996NFW,1997NFW} represented by
\begin{equation}
\label{eq:NFW}
  \rho(r) = \frac{\rho_s}{(r/r_s)(1+r/r_s)^{2}},
\end{equation}
where ${\rho_s}$ is the characteristic density and $r_s$ is the scale radius.
Here we take the values of ${\rho_s}$ and $r_s$ as 0.418 ${\rm \GeV\,cm^{-3}}$ and 16.5 ${\rm kpc}$, respectively~\cite{2012TammA&A}.

The DM density around IMBHs could be locally and dramatically enhanced by the adiabatic growth of IMBHs, forming the minispikes~\cite{1999GondoloPhRvL}. Here a non-rotating black hole (BH) is considered, and the DM profile in such a minispike is given as follows~\cite{2018LacroixApJ}:
\begin{equation}
\rho(r) = 
\begin{cases}
0 & r \leqslant 2 R_{\mathrm{S}} \\
\rho_{\mathrm{sat}} & 2 R_{\mathrm{S}} < r \leqslant R_{\mathrm{sat}} \\
\rho_{0} \left( \dfrac{r}{R_{\mathrm{sp}}} \right)^{-\gamma_{\mathrm{sp}}} & R_{\mathrm{sat}} < r \leqslant R_{\mathrm{sp}}
\end{cases},
\end{equation}
where the Schwarzschild radius is represented by $R_{\mathrm{S}}= 2G\MBH/c^2 = 2.95\  \mathrm{km}\ (\MBH/\Msolar)$, the saturation radius is given as $R_{\mathrm{sat}} = R_{\mathrm{sp}} (\rho_{\mathrm{sat}}/\rho_{0})^{-1/\gamma_{\mathrm{sp}}}$ to ensure the continuity of density, and the radius of the spike is defined as the typical radius of the BH influence $R_{\mathrm{sp}} = G \MBH\sigma_{*}^{-2} = 0.0043 \ \mathrm{pc}\ (\MBH/\Msolar)(\sigma_{*}/\mathrm{km\ s^{-1}})^{-2}$~\cite{1972PeeblesApJ,2021FortesJCAP}. The parameters $G$ and $\sigma_{*}$ are the gravitational constant and the stellar velocity dispersion, separately. We adopt the $\MBH$-$\sigma_{*}$ relation for IMBHs $\mathrm{log}(\MBH/\Msolar) = 8.13+4.02\ \mathrm{log}(\sigma_{*}/200\ \mathrm{km\ s^{-1}})$ as given in Ref.~\citep{2002TremaineApJ}.
Then, the saturation density $\rho_{\mathrm{sat}} = m_{\mathrm{\chi}}/(\sv t_{\mathrm{BH}})$ relies on the DM mass $m_{\mathrm{\chi}}$, the annihilation cross section $\sv$ and the age of BH $t_{\mathrm{BH}}$~\cite{2014LacroixPhRvD}. Here we focus on the IMBHs formed in the early universe and adopt $t_{\mathrm{BH}} \approx 10^{10}~{\rm yr}$.
All the mass inside the minispike $M_{\mathrm{sp}}$ is required to be of the order of the BH mass $\MBH$, approximately with $\rho_{0} \approx (3-\gamma_{\mathrm{sp}}) \MBH/(4 \pi R_{\mathrm{sp}}^{3})$~\cite{2018LacroixApJ}.
Considering the initial NFW distribution in minispikes, we adopt $\gamma_{\mathrm{sp}} = 7/3$ as the redistributed spike index~\cite {2005BertonePhRvD,2007FornasaPhRvD}.

\subsection{IMBHs formed in the early universe}\label{sec:IMBHs}

Here we briefly introduce two different scenarios for IMBHs formed in the early universe. 
The first scenario of IMBHs is originating from the Pop. III (or ``first") stars collapsing, called $Scenario\ PopIII$~\cite{2000SchneiderMNRAS,2001FryerApJ,2003HegerApJ,2003MarigoA&A}. The final fate of zero-metallicity Pop. III stars depends on their initial stellar masses.
Pop. III stars with the initial masses of $ 60~\Msolar - 140~\Msolar$ would collapse to black holes after the final explosion~\cite{2003MarigoA&A}.
In the mass range from roughly 140~\Msolar up to 260~\Msolar, the Pop. III star would produce a pair-instability supernova, without leaving a remnant~\cite{2001FryerApJ}.
For the high mass Pop. III star larger than 260~\Msolar, the collapse into black hole can directly occur without explosion~\cite{2003MarigoA&A}.
From above, the collapse of Pop. III stars with the mass $ 60~\Msolar \leqslant \MBH \leqslant 140~\Msolar$ and $\MBH \geqslant 260~\Msolar$ could lead to the formation of IMBHs at $\sim 100~\Msolar$~\cite{2001MadauApJ,2005BertonePhRvD}.
The second scenario of IMBHs is formed directly by the collapse of Primordial Cold Gas at high redshift, called $Scenario\ PCGas$~\cite{2001GraGnedinCQ,2003BrommApJ,2004KoushiappasMNRAS}.
The protogalactic disk could be formed in the center of early-forming halo. Due to the gas cooling and the gravitational effect, the unstable disc could eventually collapse into an IMBH with a characteristic mass of $\sim 10^5~\Msolar$~\cite{2004KoushiappasMNRAS}.

\section{Limit on the population of IMBHs}\label{sec:LPIMBHs}
\subsection{Data and anaslysis}\label{data analysis}

\begin{figure}[t] 
\centering 
\includegraphics[width=0.8\textwidth]{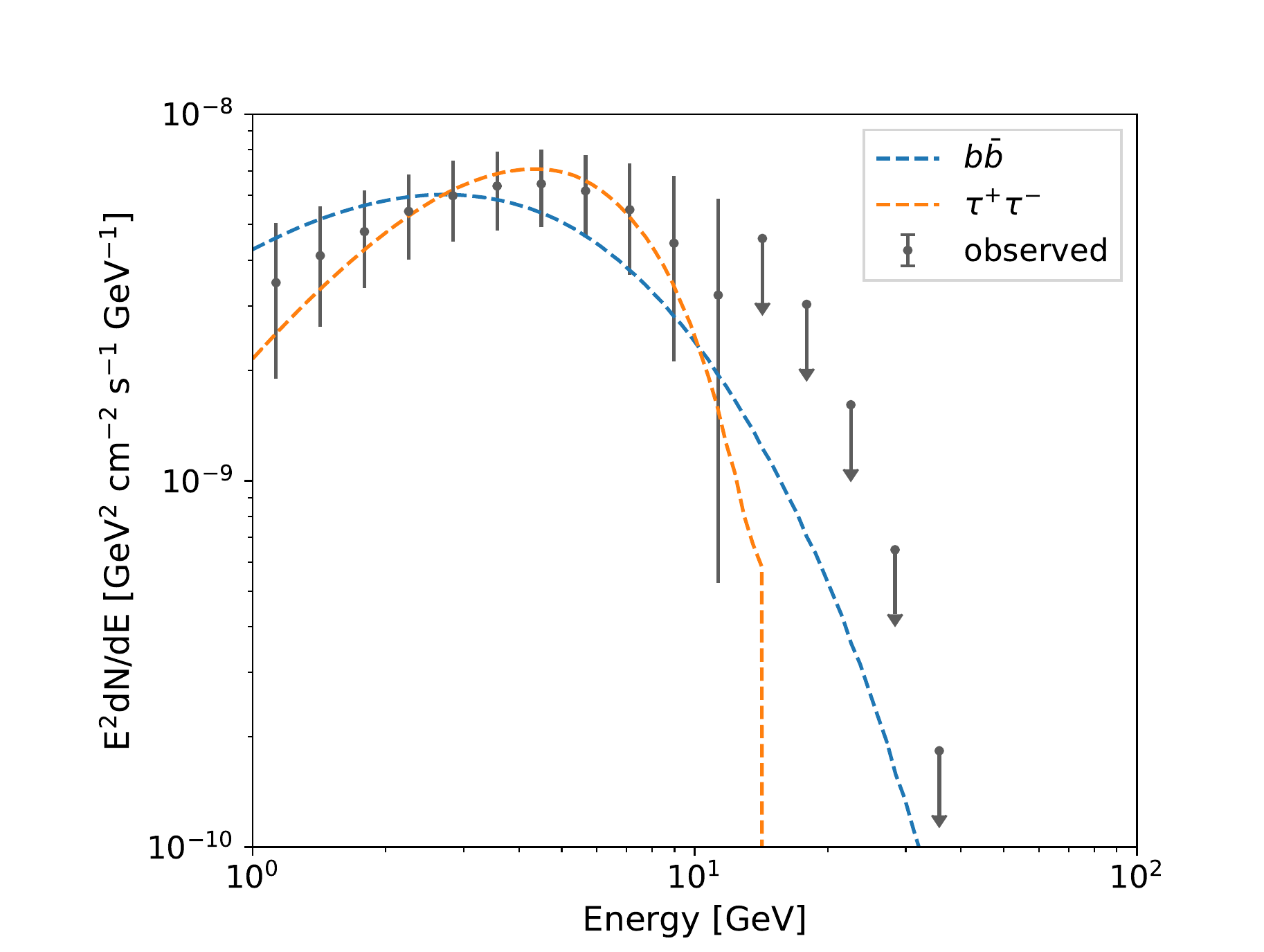}
\caption{The energy spectrum of M31SH. The grey error bars represent the measured gamma-ray spectrum of M31SH in Ref.~\cite{2019KarwinApJ}. The blue (orange) dashed line is the best fit spectrum of DM annihilation we get with the $\bb$ ($\tautau$) channel.} 
\label{Fig:sed} 
\end{figure}

The Large Area Telescope aboard the Fermi Space Telescope Mission ($Fermi$-LAT) is a current on-orbit high performance gamma-ray telescope for the photon energy range from 50 MeV to 1 TeV~\cite{2020AbdollahiApJS}.
Using nearly 7.6 years of the $Fermi$-LAT observations, Ref.~\cite{2019KarwinApJ} has carefully studied the gamma-ray emission from the M31's spherical halo with 20 evenly logarithmically spaced energy bins from 1 GeV to 100 GeV. The radial extension of the spherical halo ranges from 0.4$^{\circ}$ to 8.5$^{\circ}$ corresponding to a projected radius from 5.5 kpc to 117 kpc.
Here we directly apply the best fit energy spectrum of M31SH from Ref.~\cite{2019KarwinApJ} to study the minispikes in M31, shown as the grey error bars in the Fig.~\ref{Fig:sed}.

We perform the $least\ square$ method and the $\chi^2$ analysis in the fit procedure.
We notice that there are a few upper limits of flux in the spectrum of M31's SH which are necessary to be taken into account in the fit.
To consider upper limits, we define the $\chi^2$ by two terms as ~\cite{1986IsobeApJ,2016LyuApJ,2021KarwinPhRvD}
\begin{equation}
\label{chi1}
\chi^2 = \chi^2_{\rm cl} + \chi^2_{\rm upl}.
\end{equation}
The first term on the left side $\chi^2_{\rm cl}$ is the classic $\chi^2$ function for $n$ measurements
\begin{equation}
\label{chi2}
\chi^2_{\rm cl} = \sum_i^n \left( \frac{f_i - \hat{f}_i(\theta)}{\sigma_i} \right)^2,
\end{equation}
where $f_i$, $\sigma_i$ and $\hat{f}_i(\theta)$ denote the measured flux, uncertainty and excepted flux for the model parameters $\theta$ in the $i$th energy bin, respectively.
The second term $\chi^2_{\rm upl}$ quantifies the effect of $m$ upper limits in our fit, defined by a set of the probability ${\rm P}$ with the excepted flux $\hat{f}_j$ smaller than the upper limit $f_{{\rm upl},j}$ 
\begin{equation}
\label{chi3}
\chi^2_{\rm upl} = -2 \mathrm{ln}(\prod_{j}^{m} {\rm P}(\hat{f}_j(\theta) < f_{{\rm upl},j}))= -\sum_j^m 2 \mathrm{ln} \left( \frac{1 + \mathrm{erf} (\frac{f_{{\rm upl},j} - \hat{f}_j(\theta)}{\sqrt{2}\sigma_j})}{2}\right),%
\end{equation}
where the error function $\mathrm{erf}(x)$ is
\begin{equation}
\label{chi5}
\mathrm{erf}(x) = \frac{2}{\sqrt{\pi}}\int_0^x e^{-t^2}dt.
\end{equation}
For the observational spectrum of M31SH in Fig.~\ref{Fig:sed}, we have the number of measurements $n = 11$, the number of upper limits $m = 9$ and the total $n+m=20$.

\begin{figure}[b] 
\centering 
\includegraphics[width=0.49\textwidth]{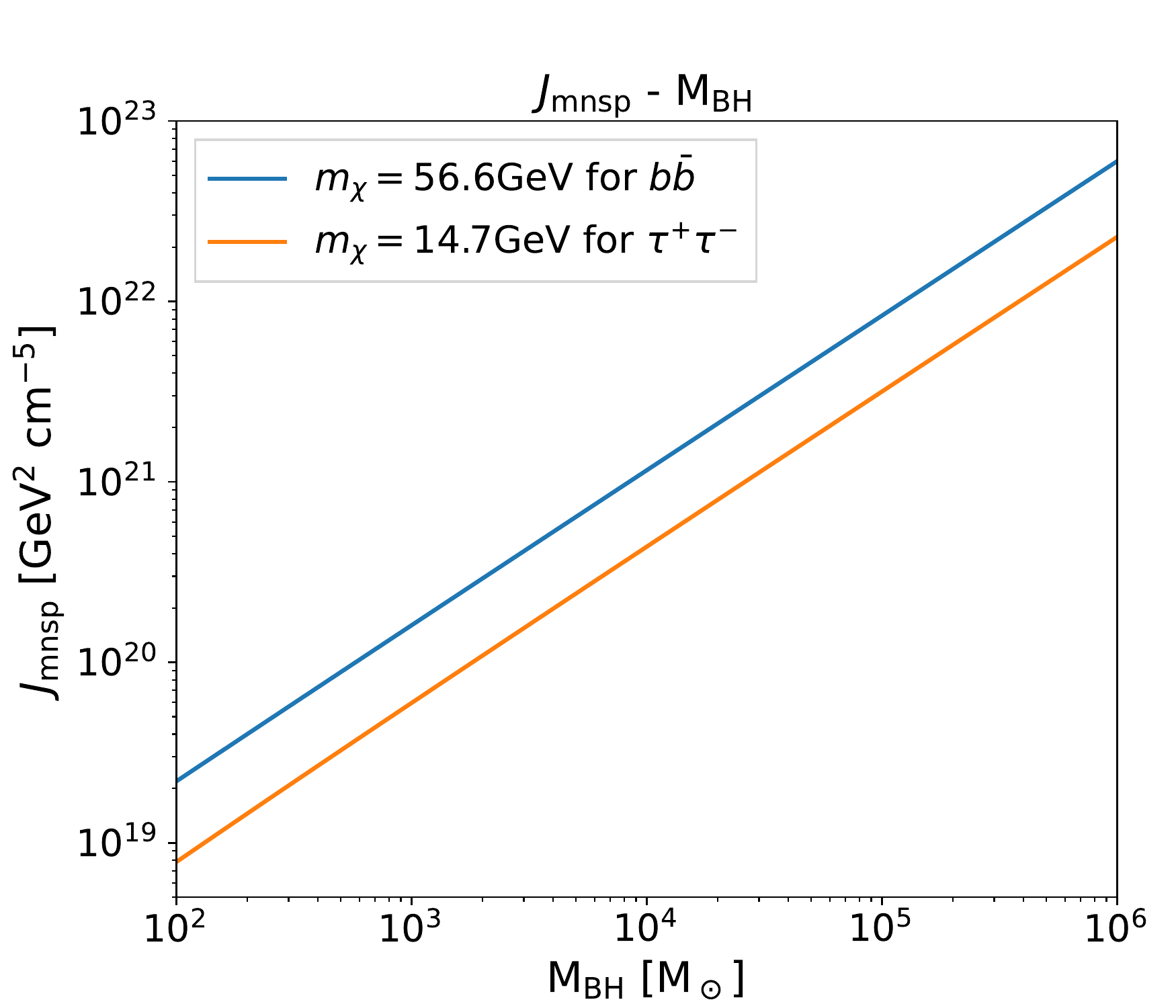}
\includegraphics[width=0.49\textwidth]{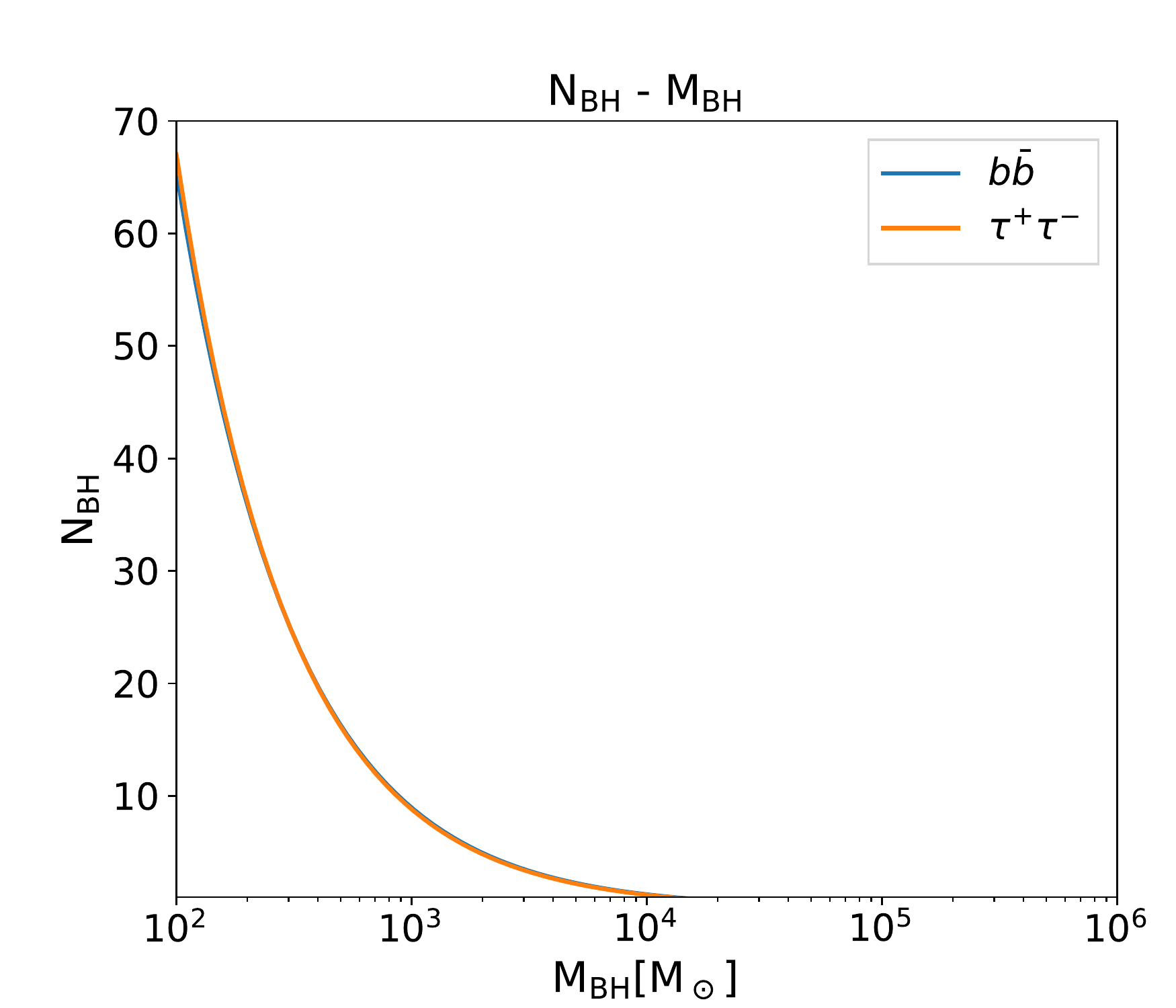}
\caption{The left panel shows the relationship between the $J_{\rm mnsp}$ for DM minispike and the IMBHs mass $\MBH$ given the best fit DM mass and the thermal relic cross section. The right panel shows limits on the number of IMBHs ${\rm N_{BH}}$ with the mass of $\MBH$ in M31SH.
For both panels, the blue and orange lines are on behalf of the $\bb$ and $\tautau$ channels, respectively.} 
\label{Fig:LPIMBHs} 
\end{figure}

\subsection{Result}\label{sec:nbh}

In this work, we plan to limit the population of IMBHs by using of the thermal relic cross section $\sv = 3 \times 10^{-26}~\svu$ for s-wave. 
For simplicity, we assume IMBHs in the M31SH have the same mass of $\MBH$.

We fit the measured gamma-ray spectrum of M31SH in the Fig.~\ref{Fig:sed} with the DM annihilation model in Eq.~(\ref{dPhide}) with two DM annihilation channels $\bb$ and $\tautau$, respectively.
The DM mass $m_\chi$ and the $\jf$ are set to be free in this fit process.
We obtain that the best fit parameters for the $\bb$ annihilation channel are $m_\chi = 56.6~\GeV$ and $\jf= 1.5\times 10^{21}~\GeV^{2}\cm^{-5}$ ($m_\chi = 14.7~\GeV$ and $\jf=5.5\times 10^{20}~\GeV^{2}\cm^{-5}$ for the $\tautau$ channel).
It is worth noting that the best fit DM mass for the $\bb$ channel is very close to that of Ref.~\cite{2021AbdughaniarXiv} ($\sim 60~\GeV$) which can explain the muon $g-2$ measurement \cite{g2}, the gamma-ray excess at the Galactic center~\cite{2017AckermannApJGCE} and the antiproton excess of AMS-02~\cite{2017CuiPhRvL,2017PhRvL.118s1102C} together.

The $\jf$ for the M31SH field can be represented as the sum of the smooth NFW component and DM minispikes around IMBHs:
\begin{equation}
\label{Jtot}
\jf = J_{\rm NFW} + {\rm N_{BH}}\times J_{\rm mnsp},
\end{equation}
where $J_{\rm NFW}=2.9\times 10^{19}~\GeV^{2}\cm^{-5}$ is integrated within the area of M31SH ($0.4^{\circ} - 8.5^{\circ}$), and the contribution from one DM minispike $J_{\rm mnsp}$ relies on the mass of IMBHs $\MBH$. When the best fit DM mass $m_\chi$ and the thermal relic cross section $\sv = 3 \times 10^{-26}~\svu$ is adopted, the relationships between the $J_{\rm mnsp}$ and the IMBHs mass $\MBH$ are given in the left panel of Fig.~\ref{Fig:LPIMBHs}.

For a set of fixed IMBH mass $\MBH$ from $10^2~\Msolar$ to $10^6~\Msolar$, we can calculate the number of IMBHs in M31SH ${\rm N_{BH}}$ from the best fit $\jf$ obtained above. The results we get for different channels ($\bb$ and $\tautau$) are very close to each other as shown in the right panel of Fig.~\ref{Fig:LPIMBHs} and we find that the number of IMBHs with masses exceeding $10^4~\Msolar$ surrounded by the DM minispikes is below one.
In the Sec.~\ref{sec:IMBHs}, we note that the IMBHs from the $Scenario\ PopIII$ have the masses of $\sim 100~\Msolar$, while the masses of the IMBHs in the $Scenario\ PCGas$ are relatively large at $\sim 10^5 ~\Msolar$.
It implies that more than 65 IMBHs with masses of $ 100 ~\Msolar$ surrounded by the minispikes are expected to locate in the M31SH field as the remnants of Pop. III stars and rule out the existence of minspikes around IMBHs in the $Scenario\ PCGas$ for both the $\bb$ and $\tautau$ channels.

\section{Discussion}\label{sec:discussion}
In the previous section, we use the standard assumption of thermal relic DM annihilation cross section to limit the population of IMBHs in M31SH. In this section, we discuss the $Scenario\ PopIII$ and $Scenario\ PCGas$ of IMBHs with the simulations to constrain the parameters of DM ($m_\chi$ and $\sv$).

Ref.~\cite{2005BertonePhRvD} has simulated the evolution of IMBHs in Milky Way galaxy starting from the early universe to the redshift z $= 0$ and constructed 200 Monte-Carlo realizations of the IMBHs population in the halo of Milky Way for each scenario, respectively.
The property of the IMBHs population in M31 is assumed to be similar to that in the Milky Way. 
We take the mass distribution of IMBHs in Ref.~\cite{2005BertonePhRvD}. 
Considering the number of IMBHs is almost proportional to the virial mass of the host galactic halo, we rescale the number of IMBHs for M31 with the masses of Milky Way and M31 in Table I of the Ref.~\cite{2007FornasaPhRvD}.
It is noticed that we only consider IMBHs in M31SH with the galactocentric distance from 5.5 kpc to 117 kpc. 
The radial distribution of IMBHs in M31 can be obtained by rescaling that of Milky Way in Fig.2 of the Ref.~\cite{2005BertonePhRvD} with the virial radius of M31 180~$\kpc$, and we find that almost 80$\%$ of IMBHs in M31 are located in M31SH for both scenarios.

In the case of $Scenario\ PopIII$, there are about ${\rm N_{BH}}=558$ IMBHs in M31SH and the mass of IMBHs is given by the delta function of $100~\Msolar$~\cite{2005BertonePhRvD}.
The total $\jf$ in the M31SH field is given as the Eq.~(\ref{Jtot}). 
While for the $Scenario\ PCGas$, we get that the number of IMBHs in M31SH is about ${\rm N_{BH}}=52$ and the distribution of IMBHs mass is predicted to obey the $log-normal$ distribution with the average mass $\MBH = 10^5~\Msolar$ and a standard deviation $\sigma_{BH} = 0.9$ as given in Ref.~\cite{2001BullockApJ,2005BertonePhRvD}. Then we generate 300 Monte-Carlo realizations for 52 IMBHs masses in M31SH.
For each realization, the total $\jf$ can be defined by
\begin{equation}
\label{JtotPCGas}
\jf = J_{\rm NFW} + {\sum_{i}^{52}} J_{\rm mnsp}(i),
\end{equation}
where $J_{\rm mnsp}(i)$ is the $\jf$ of the minispike for the $i$th IMBH.

When $\jf$ prepared, we fit the spectrum of M31SH with the DM annihilation model in Eq.~(\ref{dPhide}) using the same $\chi^2$ method of the Sec.~\ref{data analysis}.
The DM mass $m_\chi$ and annihilation section $\sv$ are  set to be free and the $\chi^2$ can be described as a function of ($m_\chi$, $\sv$).
The best-fit DM parameters for three models ($NFW$, $Scenario\ PopIII$ and $Scenario\ PCGas$) are presented in Tab.~\ref{bestfitDM}.
It is noted that we only take the $J_{\rm NFW}$ into account ($\jf=J_{\rm NFW}=2.9\times 10^{19}~\GeV^{2}\cm^{-5}$, with no minispike in M31SH) for the $NFW$ model. 
The best-fit DM parameters for the $NFW$ model are in tension with current limits from searches of dwarf spheroidal galaxies. 
As with the $Scenario\ PCGas$, we perform the same analysis with for each realizations and take the average result of all 300 realizations.
The $\chi^2$ values with a series of ($m_\chi$, $\sv$) are shown as colour maps in Fig.~\ref{fig:dmupl} of the Appendix. 

The best fit annihilation cross sections $\sv$ for $Scenario\ PopIII$ and $Scenario\ PCGas$ are significantly less than $\rm 3 \times 10^{-26}~cm^3~s^{-1}$ which is the DM s-wave cross section yielding the correct relic density.
It may indicate the possibility of the p-wave or d-wave DM annihilation where the cross sections rely on the velocity of DM particles ($v$)~\cite{2014BerlinPhRvD}. 
If the velocity of DM particles in the minispikes around IMBHs is about several hundreds to several thousands kilometers per second, the current annihilation cross section $\sv$ for p-wave is $ \sim (v/c)^2\sim 10^{-29}~\svu$ and for d-wave is $ \sim (v/c)^4\sim 10^{-36}~\svu$, which are exactly the best fit $\sv$ for $Scenario\ PopIII$ and $Scenario\ PCGas$ in Tab.~\ref{bestfitDM}, respectively.

Then we derive the exclusion on DM parameters.
For a set of fixed DM masses $m_\chi$ in the range $1-200~\GeV$, we calculate the $\chi^2$ values in the Eq.~(\ref{chi1}) with the different annihilation cross sections $\sv$. The 95$\%$ confidence level upper limit of $\sv$ is obtained when the $\chi^2$ value is larger by 2.7 than the least value for each $m_\chi$. The upper limits of three models are represented as the black lines in the Fig.~\ref{fig:dmupl}.

\begin{table*}[!t]
\caption{The best-fit DM parameters.}
\begin{tabular}{ccccc}
\hline
\hline
Annihilation Channel  & Model & $m_\chi(\GeV) $  &  $\sv (\svu)$ & {Reduced $\chi^2$}\\
\hline
$\bb$ & $NFW$  & 56.6 $\pm \ 8.5$ & $(1.25 \pm 0.26) \times10^{-24}$  & 0.47 \\
$\bb$ & $Scenario\ PopIII$  & 56.6 $\pm \ 8.5$  & $(1.56 \pm 0.48) \times10^{-29}$  & 0.47\\
$\bb$ & $Scenario\ PCGas$  & 56.6 $\pm \ 8.5$ & $(1.46 \pm 0.35) \times10^{-36}$ & 0.47 \\
\hline
$\tautau$ & $NFW$ & 14.7 $\pm \ 1.6$ & $(5.75 \pm 0.81)\times10^{-25}$  & 0.21 \\
$\tautau$ & $Scenario\ PopIII$ & 14.7 $\pm \ 1.6$ & $(1.51 \pm 0.45)\times10^{-29}$  & 0.21\\
$\tautau$ & $Scenario\ PCGas$ & 14.7 $\pm \ 1.6$ & $(1.18 \pm 0.31) \times10^{-36}$ & 0.21\\
\hline
\hline
\end{tabular}
\label{bestfitDM}
\end{table*}

\section{Summary}\label{sec:summary}

In this work, we investigate the DM explanation of gamma-ray excess from the M31SH region using the assumption of minispikes around IMBHs.
First, we study the population of IMBHs in M31SH with the thermal relic cross section.
We fit the M31SH’s spectrum data with the DM annihilation model. The best-fit $\jf$ of M31SH we obtain is larger than the smooth NFW component $J_{\rm NFW}$, which could be contributed by the DM minispikes around IMBHs located in the M31BH field.
For the simplified assumption of IMBHs with the same mass, we infer the relationship between the number of IMBHs in M31SH $N_{\rm BH}$ and the IMBHs mass $\MBH$ shown in the right panel of Fig.~\ref{Fig:LPIMBHs}.
For $\sim 100~\Msolar$ IMBHs in the $Scenario\ PopIII$, we find that above 65 DM minspikes is needed to yield the observed spectrum of M31SH for both annihilation channels $\bb$ and $\tautau$.
And IMBHs with the masses above $10^4~\Msolar$ in the $Scenario\ PCGas$ surrounded by the DM minspikes can be excluded by the adoption of the thermal relic cross section.

However the nature of WIMPs is still unclear.
Then we discuss the two different scenarios for IMBHs ($Scenario\ PopIII$ and $Scenario\ PCGas$ ) with simulations to study the properties of DM.
We rescale the IMBHs population of the Milky Way halo in the Ref.~\cite{2005BertonePhRvD} for M31 and calculate the total $\jf$ of M31SH for each scenario individually.
Then the spectrum of M31SH is refitted with the DM annihilation model to optimize the DM parameters ($m_\chi$ and $\sv$). The best-fit DM masses and annihilation sections $\sv$ are listed in the Tab.~\ref{bestfitDM} and the upper limits are derived in the Appendix.

\acknowledgments
We thank Christopher M. Karwin for providing us with the spectrum data of M31SH, and Qiang Yuan, Yue-Lin Sming Tsai, Xiaoyuan Huang, Zhanfang Chen and Guan-Wen Yuan for valuable comments and discussions. We use the {\tt NumPy}~\cite{NumPy}, {\tt SciPy}~\cite{SciPy}, {\tt Matplotlib}~\cite{Matplotlib} and {\tt iminuit}\footnote{\url{https://github.com/scikit-hep/iminuit}} packages in our analysis.
This work is supported by the National Natural Science Foundation of China (No. U1738210, No. 12003069, No. 11773075, No. 12003074 and No. 12047560), the National Key Program for Research and Development (No. 2016YFA0400200) and the Entrepreneurship and Innovation Program of Jiangsu Province.

\bibliography{references}

\bibliographystyle{JHEP}

\clearpage
\section*{A: Constraints on the DM parameters}\label{sec:Aupl}
 
\begin{figure}[htbp] 
\centering 
\includegraphics[width=0.49\textwidth]{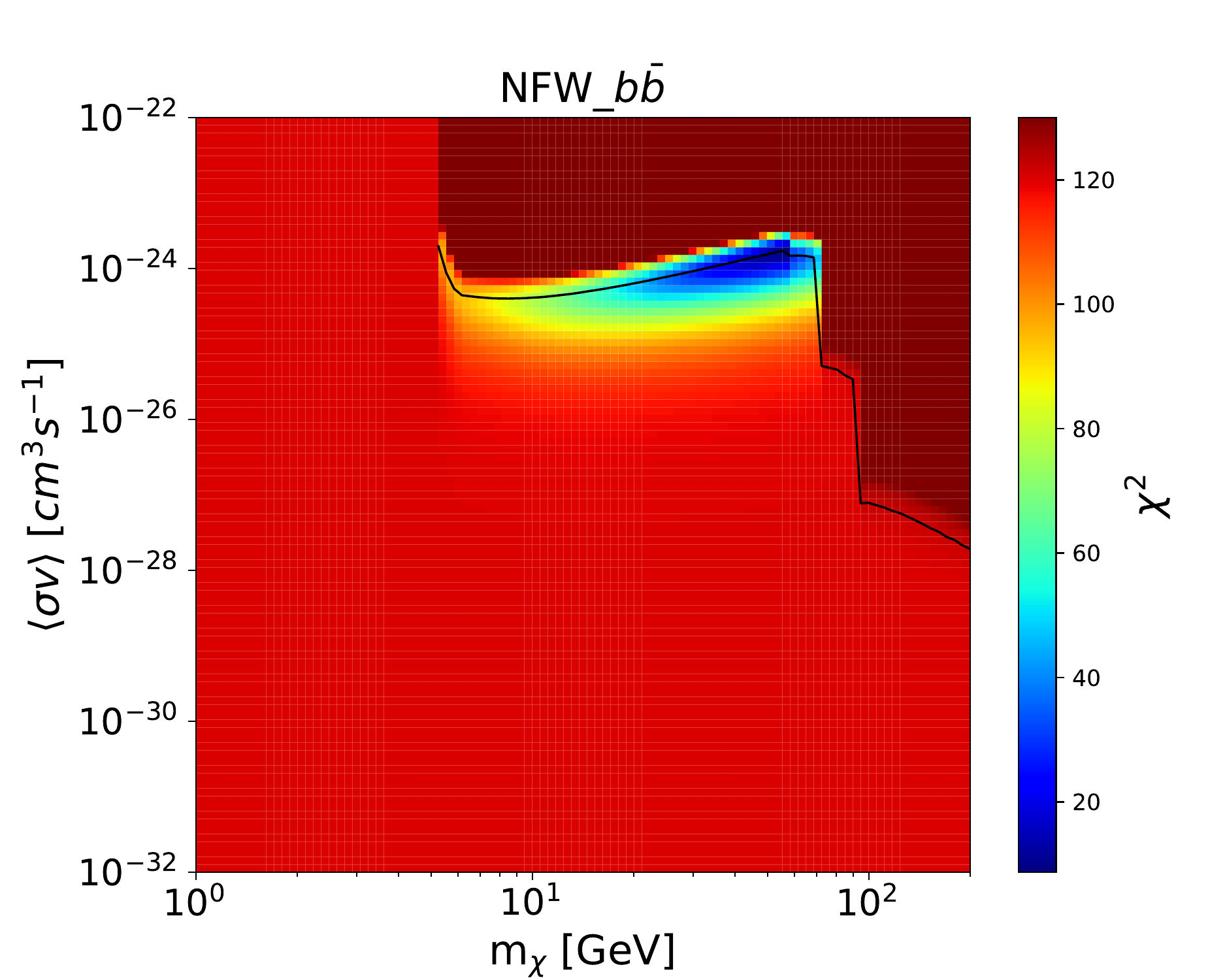}
\includegraphics[width=0.49\textwidth]{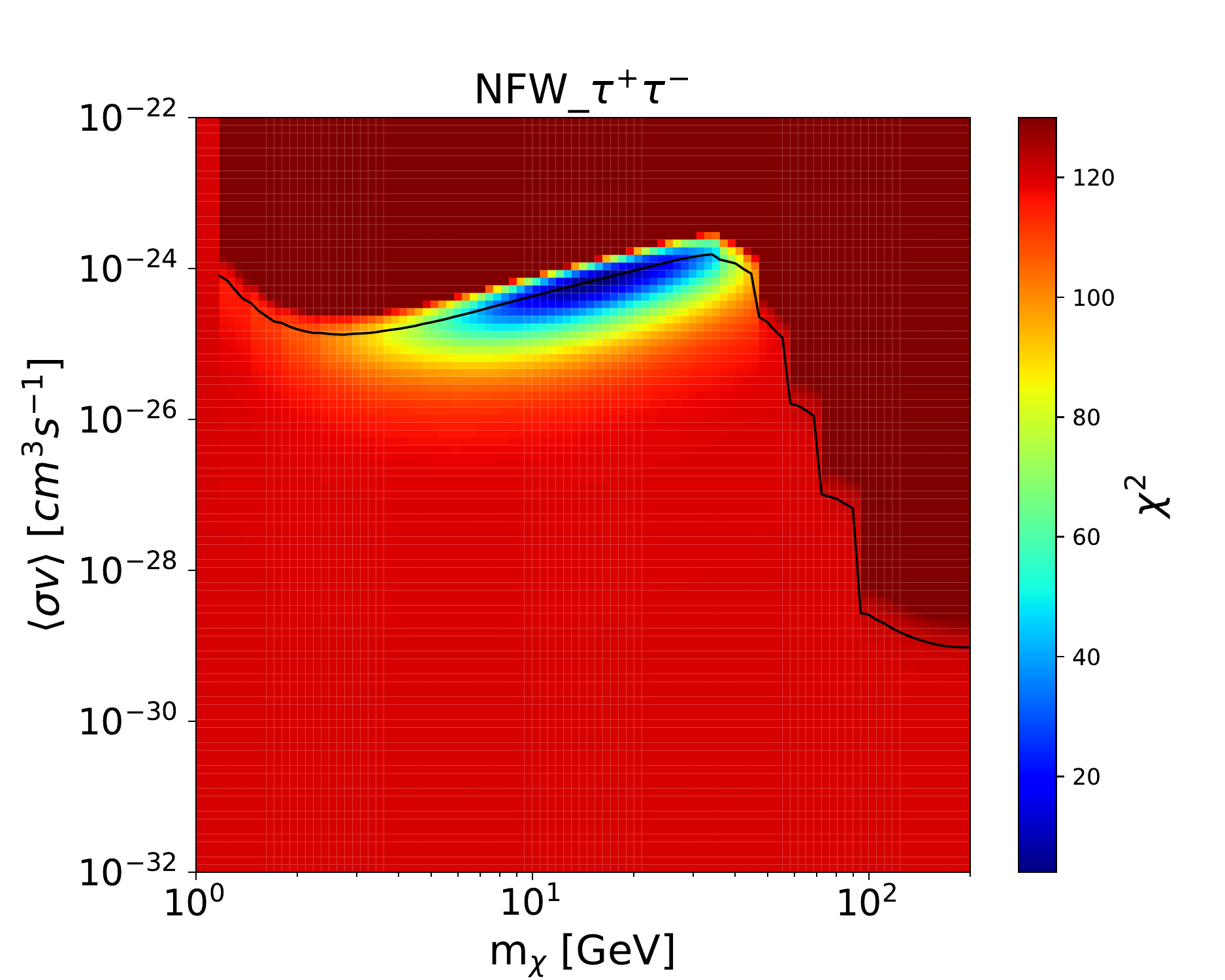}
\includegraphics[width=0.49\textwidth]{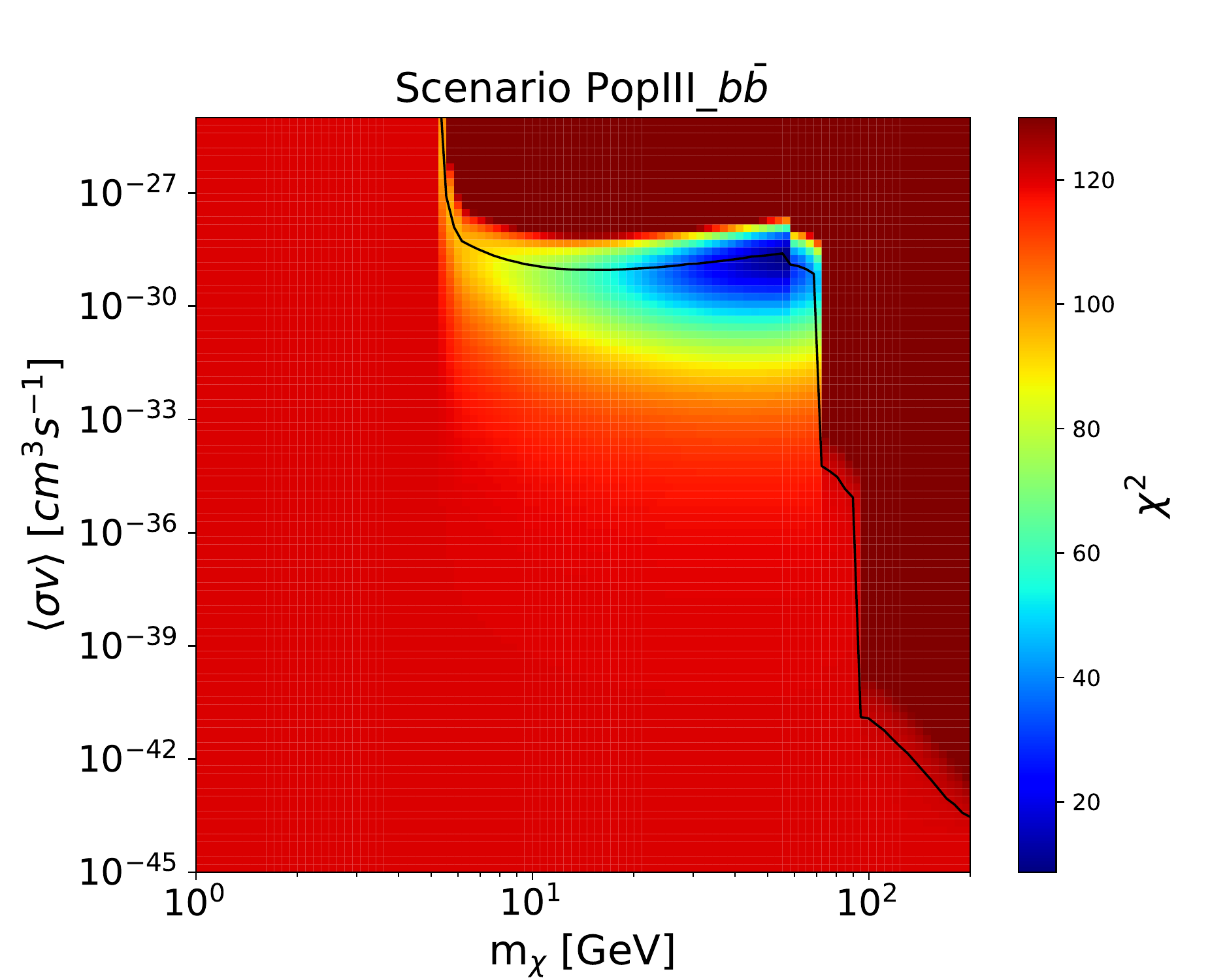}
\includegraphics[width=0.49\textwidth]{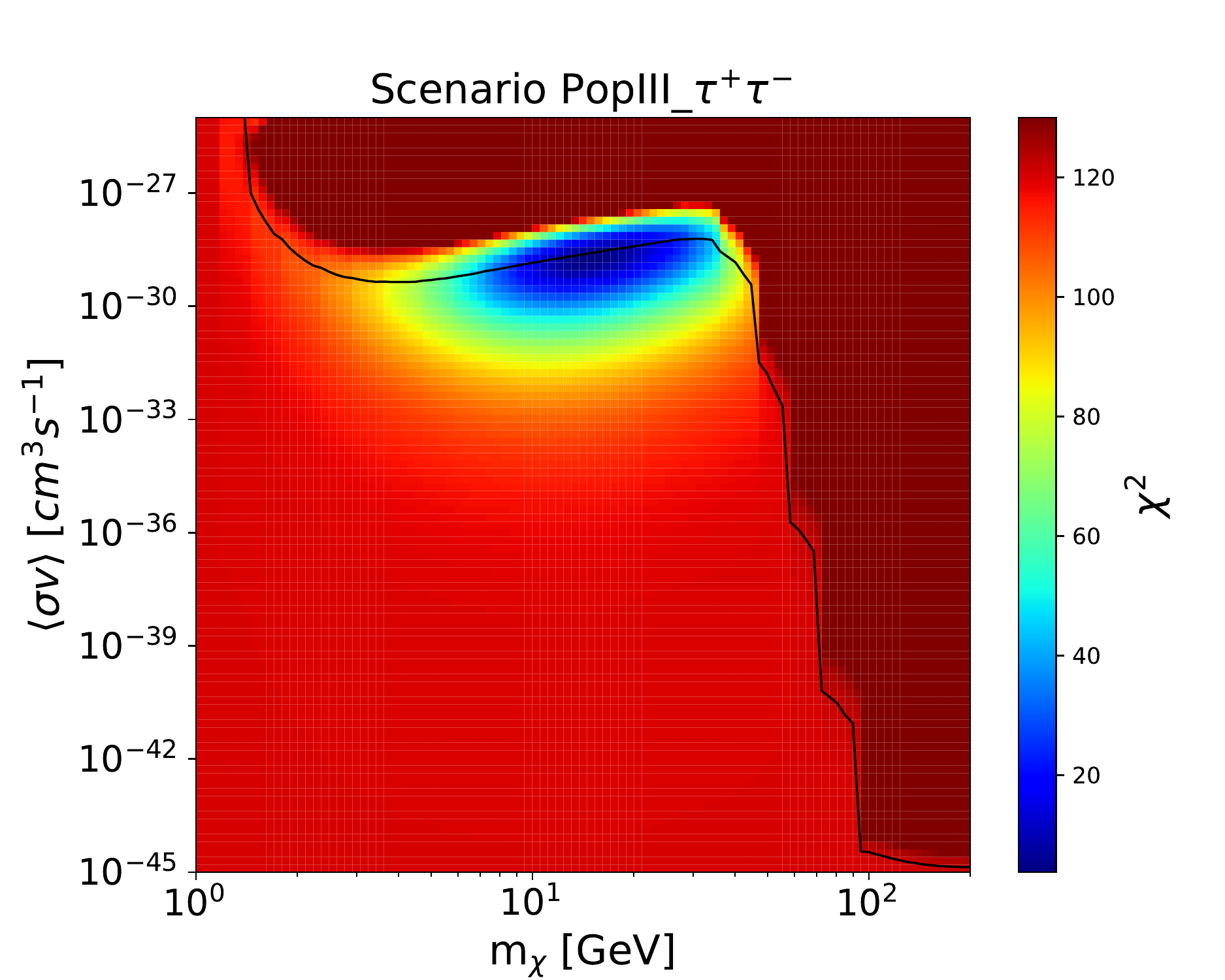}
\includegraphics[width=0.49\textwidth]{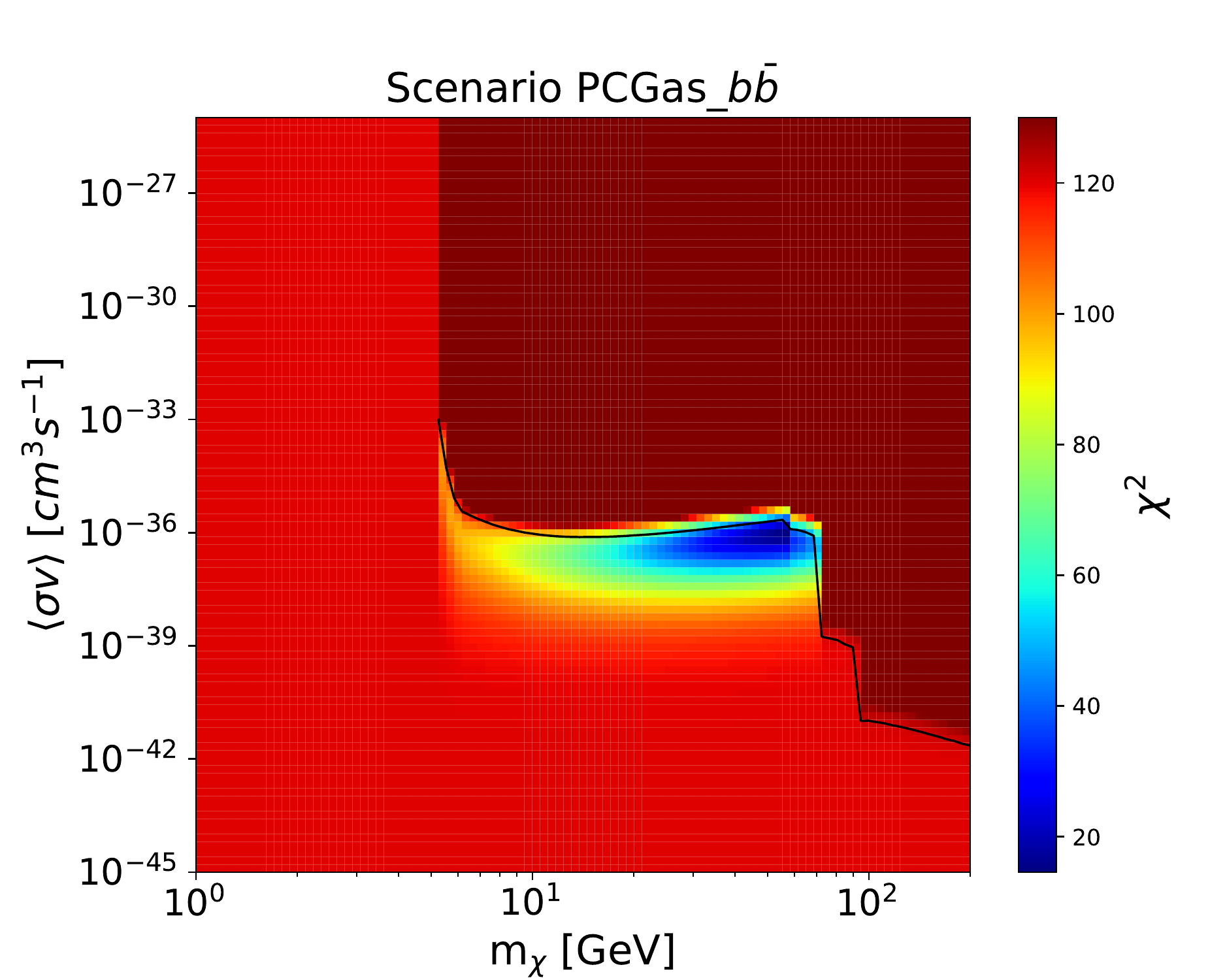}
\includegraphics[width=0.49\textwidth]{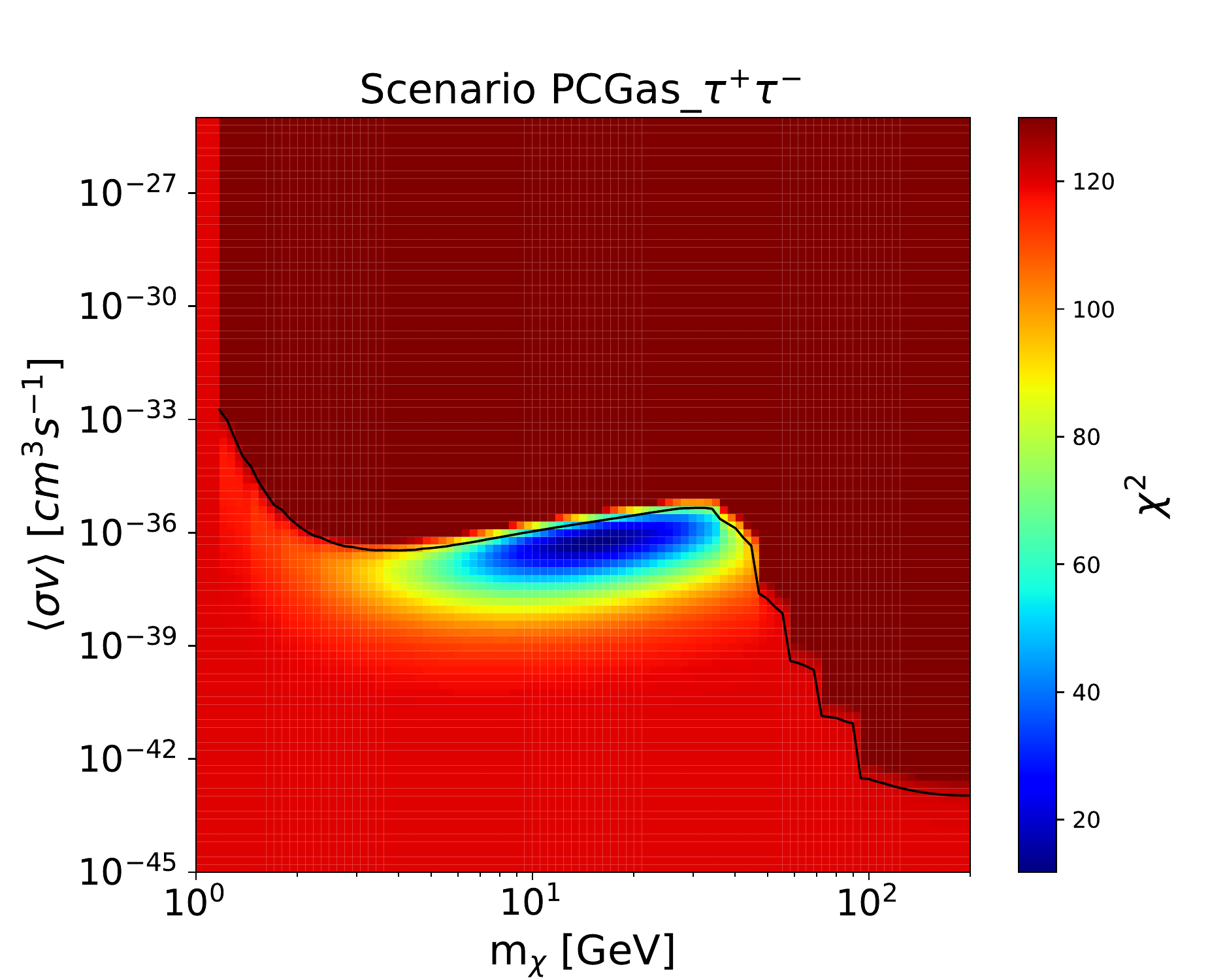}
\caption{ The $\chi^2$ value as a function of the DM mass $m_\chi$ and annihilation cross section $\sv$ for three models ($NFW$, $Scenario\ PopIII$ and $Scenario\ PCGas$) with the $\bb$ and $\tautau$ channels, separately. The black lines are the $95\%$ confidence level upper limits of $\sv$.} 
\label{fig:dmupl}
\end{figure}

\end{document}